\documentclass[pmlr]{jmlr}

\RequirePackage{graphicx}
 \usepackage{booktabs}
\usepackage{longtable}
\usepackage{bm}
\usepackage{pifont}
\newcommand{\cmark}{\ding{51}}%
\newcommand{\xmark}{\ding{55}}%

\makeatletter
\def\set@curr@file#1{\def\@curr@file{#1}} 
\makeatother
\usepackage[load-configurations=version-1]{siunitx} 

\theorembodyfont{\upshape}
\theoremheaderfont{\scshape}
\theorempostheader{:}
\theoremsep{\newline}

\jmlrvolume{298}
\jmlryear{2025}
\jmlrworkshop{Machine Learning for Healthcare}

\title[FIVA: Federated Inverse Variance Averaging for Universal CT Segmentation]{FIVA: Federated Inverse Variance Averaging for Universal CT Segmentation with Uncertainty Estimation }

\author{\Name{Asim Ukaye  \textsuperscript{1}} \Email{asim.ukaye@mbzuai.ac.ae} \\
\Name{Numan Saeed  \textsuperscript{1}} \Email{numan.saeed@mbzuai.ac.ae} \\
\Name{Karthik Nandakumar \textsuperscript{1,2}} \Email{nandakum@msu.edu}   \\
\\
\addr \textsuperscript{1} Department of Computer Vision, Mohamed bin Zayed University of Artificial Intelligence \\
\textsuperscript{2} \addr Department of Computer Science and Engineering, Michigan State University
}

\begin{document}

\maketitle

\begin{abstract}
Different CT segmentation datasets are typically obtained from different scanners under different capture settings and often provide segmentation labels for a limited and often disjoint set of organs. 
Using these heterogeneous data effectively while preserving patient privacy can be challenging. This work presents a novel federated learning approach to achieve universal segmentation across diverse abdominal CT datasets by utilizing model uncertainty for aggregation and predictive uncertainty for inference. 
Our approach leverages the inherent noise in stochastic mini-batch gradient descent to estimate a distribution over the model weights to provide an on-the-go uncertainty over the model parameters at the client level. The parameters are then aggregated at the server using the additional uncertainty information using a Bayesian-inspired inverse-variance aggregation scheme. 
Furthermore, the proposed method quantifies prediction uncertainty by propagating the uncertainty from the model weights, providing confidence measures essential for clinical decision-making. In line with recent work shown, predictive uncertainty is utilized in the inference stage to improve predictive performance.
Experimental evaluations demonstrate the effectiveness of this approach in improving both the quality of federated aggregation and uncertainty-weighted inference compared to previously established baselines. The code for this work is made available at: \url{https://github.com/asimukaye/fiva} 

\end{abstract}

\section{Introduction}

Medical image segmentation plays a crucial role in clinical diagnostics, with applications ranging from tumor detection to organ delineation. 
While deep learning models have achieved remarkable success in this domain, data privacy regulations restrict patient data sharing, thereby limiting their full potential.
Federated Learning (FL) enables collaborative model training across distributed datasets while preserving data privacy (\cite{mcmahanCommunicationEfficientLearningDeep2017}).
In a typical horizontal FL setting, each client trains a local model on its private dataset and sends the model to a central server. The server aggregates these models to form a global model, which is then redistributed to the clients for the next training round.

However, heterogeneity in client data due to differences in data quantity, quality, or acquisition equipment is known to degrade FL performance (\cite{liFederatedOptimizationHeterogeneous2020a}). Real-world settings such as medical imaging naturally exhibit this data heterogeneity due to the diversity of patients (e.g., region, ethnicity) and differences in imaging equipment, data collection, and labeling practices at different healthcare centers. This makes the direct application of FL algorithms to medical imaging, and healthcare in general, an open problem (\cite{rauniyarFederatedLearningMedical2024}).

\begin{figure}[htbp]
\centering
\includegraphics[width=0.98\textwidth]{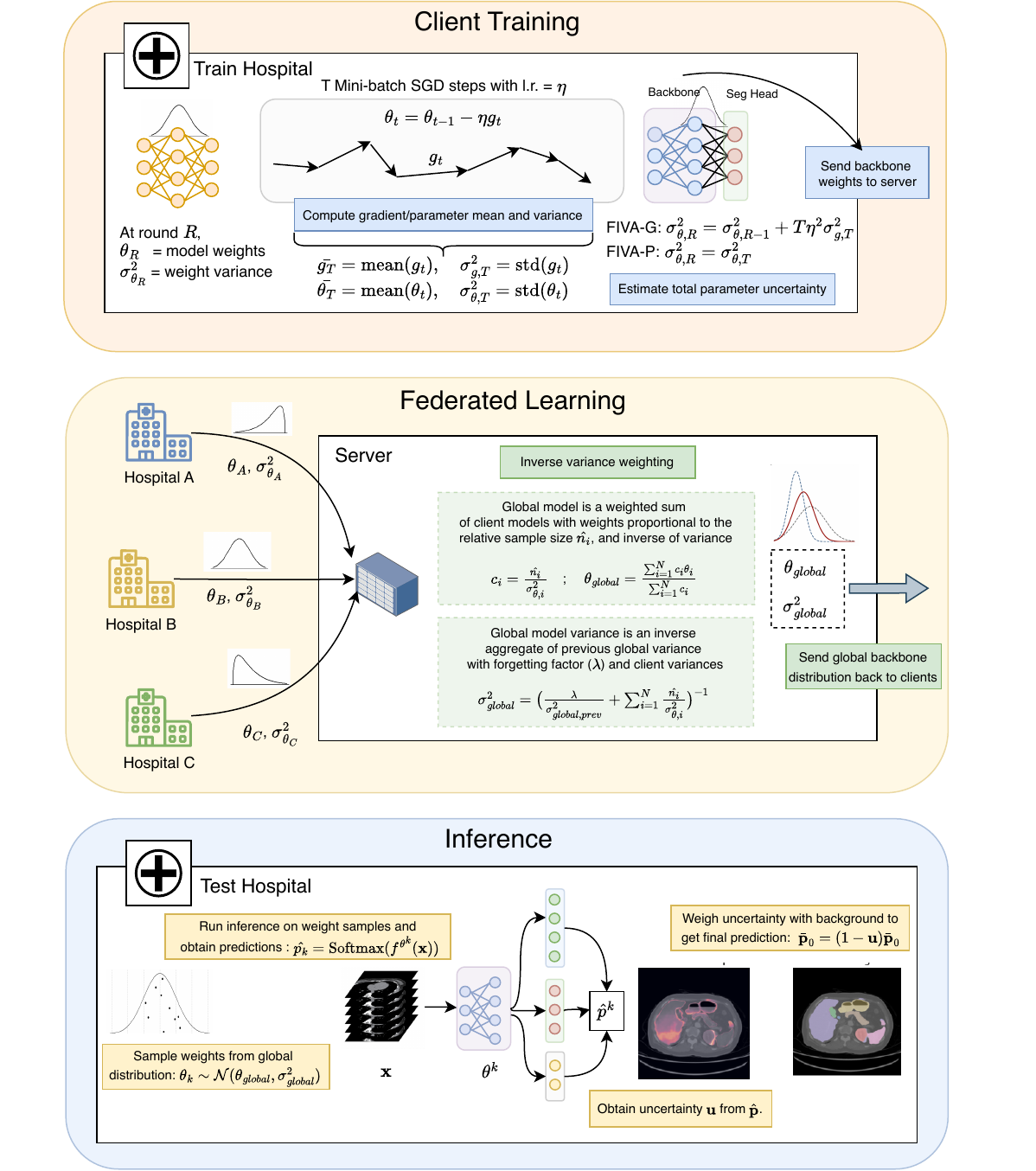}
\caption{
Overview of the proposed FIVA approach. Each client induces a distribution over model parameters during local training. These parameter distributions are aggregated on the server using inverse-variance weighting to form a global distribution. At inference time, the test client samples model parameters from the global distribution to obtain predictive uncertainty. Final predictions are computed by averaging stochastic forward passes and reweighting the background class using total uncertainty, following \cite{tolleFUNAvgFederatedUncertainty2024}.
}
\label{mainfig}
\end{figure}

\noindent On the other hand, uncertainty estimation in medical image segmentation is paramount due to its potential to enhance the reliability and interpretability of automated diagnostic tools. \cite{kendallWhatUncertaintiesWe2017} formulated uncertainty estimation in the context of deep learning to differentiate the impact of irreducible, data-dependent (aleatoric) uncertainty and reducible, model-dependent (epistemic) uncertainty. In medical image segmentation, capturing epistemic uncertainty helps to identify ambiguous regions and potential out-of-distribution samples (\cite{jalalEvaluatingUncertaintyQuantification2024}). This motivates the need to develop methods that enable federated training of medical image segmentation models while providing reliable uncertainty estimates of predictions.

\noindent Recently, \cite{tolleFUNAvgFederatedUncertainty2024} proposed a novel approach to federated learning for medical image segmentation (FUNAvg) that leverages predictive uncertainty at the inference stage. By performing uncertainty-weighted averaging of output channels, FUNAvg improves predictive performance.
We build on FUNAvg and draw inspiration from statistical meta-analysis such as inverse variance weighting for aggregating random variables (\cite{hartungStatisticalMetaanalysisApplications2008}, \cite{borensteinIntroductionMetaanalysis2013}). Based on this, we propose a novel strategy: \textbf{FIVA}: \underline{F}ederated \underline{I}nverse \underline{V}ariance \underline{A}veraging, that embeds uncertainty in the model and effectively utilizes it to improve aggregation during the federated learning step.


\noindent In this work, we propose using parameter uncertainty estimation to enhance federated learning for abdominal CT image segmentation. Each client estimates the mean and variance of the model parameters during local training by tracking the gradients and parameters at each iteration of mini-batch stochastic gradient descent. The clients then send these model parameters and their variances to the server. The central server aggregates these parameters, considering both their mean values and associated uncertainties. The resulting global model represents a posterior distribution over parameters, allowing inference on test clients by sampling from this distribution. We illustrate this approach in Figure \ref{mainfig}.

Our key contributions are as follows. 
\begin{itemize}
    \item Leverage the stochasticity in mini-batch SGD to estimate a distribution over model parameters during client training. This enables each client to capture epistemic uncertainty locally efficiently.
    \item Introduce a Bayesian aggregation framework compatible with existing FL pipelines that utilizes this parameter distribution to improve both server-side aggregation and predictive uncertainty estimation during inference. 
    \item Demonstrate improved performance over standard baselines through both quantitative and qualitative evaluations. We benchmark our method on abdominal CT segmentation tasks and show its robustness under varying data distributions.
\end{itemize}

\noindent \textbf{Generalizable insights and clinical significance.}
This approach can significantly enhance diagnostic confidence through uncertainty-aware predictions. By quantifying uncertainty, healthcare professionals can better assess the reliability of segmentation predictions and identify cases that require expert review. Federated learning allows for privacy-preserving collaboration across institutions, improving model generalization across diverse patient populations without centralized data collection.

\section{Related Work}

\subsection{Universal Segmentation}
Universal Segmentation aims to train models that generalize across datasets with diverse organ coverage, label availability, and acquisition settings.
Recently, \cite{butoi2023universeg} introduced \textit{UniverSeg}, a model capable of adapting to unseen medical segmentation tasks without requiring additional training or fine-tuning.
\cite{gaoTrainingMedicalResident2024} propose \textit{Hermes}, a context-prior learning approach that integrates task and modality priors into the segmentation process. Another line of work focuses on creating highly optimized end-to-end pipelines to enable generalization across diverse medical imaging tasks. For example,  \cite{isenseeNnUNetSelfconfiguringMethod2021} introduced \textit{nnU-Net}, a self-configuring pipeline built on the U-Net architecture (\cite{ronneberger2015u}) that is adaptable to various segmentation tasks. \cite{wasserthalTotalSegmentatorRobustSegmentation2023} propose \textit{TotalSegmentator} which can segment 104 anatomical structures covering organs, bones, muscles, and major vessels. Other works such as \textit{Multi-Talent}, (\cite{ulrichMultiTalentMultiDatasetApproach2023}), and \textit{Med3D} (\cite{chen2019med3d}) attempt to learn a shared representation between datasets while preserving label-specific heads.

\subsection{Federated Learning in Medical Imaging}
Federated Learning (FL) has emerged as a popular paradigm for training models on decentralized medical data while preserving patient privacy. 
\cite{zhangPersonalizedFederatedLearning2022} propose \textit{pFedBayes} that performs personalized federated learning by treating both client and server neural networks as Bayesian Neural Networks. In the context of medical imaging, \cite{xuFederatedMultiOrganSegmentation2023} introduces \textit{Fed-MENU}, which utilizes sub-networks that specialize in extracting features for specific organs. \cite{bernecker2022fednorm} address variations in imaging modalities and scanner types in their work \textit{FedNORM} for liver segmentation in a federated learning setting. \cite{jiangFairFederatedMedical2023a} propose \textit{FedCE} that addresses fairness in medical image segmentation.
They estimated client contributions in an FL setting using gradient updates and server-side validation. \cite{asokan2024federated} integrate Low-Rank Adapters into the \textit{Segment Anything Model (SAM)} architecture to enable parameter-efficient fine-tuning of of 3D medical image segmentation under an FL setting. \cite{xu2024federated} introduce \textit{FedCross} to train the global model in a round-robin manner eliminating the need for aggregation. They also show a novel way to estimate the predictive uncertainty for image segmentation using their approach. 


\subsection{Uncertainty Estimation in Medical Imaging}
Uncertainty quantification enhances model interpretability and safety in medical AI. \cite{gal2016dropout} show that applying Monte Carlo Dropout (MCD) at test time enables estimation of epistemic uncertainty in deep learning models. \cite{lakshminarayananSimpleScalablePredictive2017a} similarly show that deep ensembles provide well-calibrated predictive uncertainty estimates. \cite{baumgartnerPHiSegCapturingUncertainty2019} introduce \textit{PHiSeg}, which uses variational autoencoders to model segmentation uncertainty for medical images. \cite{zhangUncertaintyQuantificationFederated2023} study various uncertainty quantification techniques within federated learning frameworks and show how heterogeneous data affects the calibration and reliability of uncertainty estimates. \cite{judgeCRISPReliableUncertainty2022}
use contrastive learning and anatomical priors to generate well-calibrated and anatomically consistent uncertainty maps for medical image segmentation.
 \cite{koutsoubis2024privacy} provided a comprehensive review of federated learning and uncertainty estimation methods specific to medical imaging.

\section{Methodology}

We follow a training approach similar to the methodology used for multi-task segmentation as described in \cite{chen2019med3d}, \cite{ulrichMultiTalentMultiDatasetApproach2023}, and \cite{tolleFUNAvgFederatedUncertainty2024}. We retain a common model backbone for all clients, and allocate a separate segmentation head to each client, with output channels corresponding to the labels present in their dataset. We run the training phase of the experiments in three settings. 1) In the \textbf{centralized} setting, the common backbone is jointly trained using the data from all clients. For each batch, the loss is propagated through the segmentation head corresponding to the dataset the batch was drawn from. This setting typically serves as an upper bound for model performance. 2) In the \textbf{federated} setting, each client locally trains the common backbone along with its own segmentation head, and shares the updated backbone with the central server at the end of each communication round. The server aggregates the client models with a chosen aggregation scheme and sends the global backbone to the clients. 3) In the \textbf{standalone} setting, each client trains a model independently on its own dataset. This setting is used to assess how well a client-specific model generalizes to data from other clients.

\noindent Inference is primarily performed by averaging the logits obtained from each segmentation head. However, one of the key insights from FUNAvg is that predictive performance can be significantly improved by reweighting the background channel of the predicted logits using an uncertainty estimate. This adjustment accounts for the tendency to overestimate the background class during aggregation. They argue that aggregation from multiple heads can overestimate the background class, since the target class may not be present in all segmentation heads. They used MC Dropout as their predictive uncertainty estimation technique. In our work, we retain this uncertainty-weighted averaging scheme. However, we employ a different method for quantifying uncertainty, as described in Section \ref{sec:inference}.

\noindent We propose novel enhancements to the standard federated learning pipeline at three stages: a) local client training, b) server aggregation, and c) client inference. In the local client training stage, we introduce an online variance estimation technique for constructing the client parameter distribution.
This distribution is sent to the server, which uses an inverse variance aggregation scheme to obtain the global model parameters.
Finally, during inference, we employ a sampling strategy to estimate the predictive uncertainty over the test samples. 
To describe the problem setup, we define the standard variables used in federated learning. Let $N$ be the total number of participating clients and $R$  the total number of federated rounds. Each client performs local training for $T$ mini-batch SGD steps. Let $n_i$ denote the number of data samples held by client $i$.

\subsection{Parameter Distribution Estimation}
 Each client trains its local model using mini-batch stochastic gradient descent (SGD) and estimates parameter uncertainty by tracking the parameter or gradient statistics across SGD iterations. Let $t$ represent the current mini-batch SGD step, $\bm{\theta}_t$ the model parameters at step $t$, and $\bm{g}_t$ the corresponding gradient. The client computes the parameter update using the standard mini-batch SGD update with $\eta$ as the learning rate.
\begin{equation}
    \bm{\theta}_t = \bm{\theta}_{t-1} - \eta \bm{g}_t
\end{equation}
\noindent We propose two different approaches for estimating the parameter uncertainty:

\noindent \textbf{Gradient-based estimation.} In this approach, we estimate the parameter uncertainty by first estimating the gradient variance and then accumulating them to construct the parameter variance. To reduce the memory overhead from storing all $T$ gradient vectors, we compute the gradient mean and variance in an online fashion using Welford’s algorithm (\cite{Welford01081962}), as shown below.
\begin{equation}
\begin{split}
    \bm{\bar{g}}_t &= \bm{\bar{g}}_{t-1} + \frac{\bm{g}_t - \bm{\bar{g}}_{t-1}}{t} \\
    \bm{M}_{2,t}   &= \bm{M}_{2,t-1} + (\bm{g}_t - \bm{\bar{g}}_{t-1}) (\bm{g}_t - \bm{\bar{g}}_t) \\
    \bm{\sigma}_{g,T}^2 &= \frac{\bm{M}_{2,T}}{T} \\
\end{split}
\end{equation}

\noindent $\bm{M}_{2,t}$ is the intermediate sum of squared errors used to compute the gradient variance at the end of $T$ steps, $\bm{\sigma}_{g,T}^2$. 
The estimated gradient variance is added to the previous estimate of the total parameter variance $\bm{\sigma}_{\theta, 0}^2$ to get the new estimated parameter variance:
\begin{equation}
    \bm{\sigma}_{\theta,T}^2  = \bm{\sigma}_{\theta,0}^2  + T\eta^2\bm{\sigma}_{g,T}^2 
\end{equation}
The above equation makes a simplifying assumption of independence of gradient updates after each round to avoid computing the full gradient covariance. We refer to this variant of the algorithm as `FIVA-G' in the upcoming sections.

\noindent \textbf{Parameter-based estimation.} In this variant, we directly estimate the parameter variance by tracking the running parameter mean and variance using Welford's algorithm as shown below.
\begin{equation}
\begin{split}
    \bm{\bar{\theta}}_t &= \bm{\bar{\theta}}_{t-1} + \frac{\bm{\theta}_t - \bm{\bar{\theta}}_{t-1}}{t} \\
    \bm{P}_{2,t}   &= \bm{P}_{2,t-1} + (\bm{\theta}_t - \bm{\bar{\theta}}_{t-1}) (\bm{\theta}_t - \bm{\bar{\theta}}_t) \\
    \bm{\sigma}_{\theta,T}^2 &= \frac{\bm{P}_{2,T}}{T} \\
\end{split}
\end{equation}
$\bm{P}_{2,t}$ is the sum of squared errors for the parameters similar to $\bm{M}_{2,t}$. 
We refer to the parameter-based estimation variant of the overall algorithm as `FIVA-P' in the upcoming sections.

\noindent The model parameters obtained after $T$ mini-batch SGD steps represent the parameters for client $i$ in global federation round $r$, denoted as $\bm{\theta}_{i,r} = \bm{\theta}_T$.
Similarly, the parameter variance for client $i$ at round $r$ is denoted as $\bm{\sigma}_{\theta,i,r}^2 = \bm{\sigma}_{\theta,T}^2 $  . 
For readability, we drop the $\theta$ subscript and represent the parameter variance for client $i$ in round $r$ as $\bm{\sigma}_{i,r}^2$ in the following sections.

\subsection{Inverse Variance Aggregation}
For a given federation round $r$, each client $i$ transmits its parameter update and variance ($\bm{\theta}_{i,r}, \bm{\sigma}_{i,r}^2$) to the central server. The server then aggregates the received parameters using a weighted averaging scheme to obtain the global parameter vector and its variance, denoted by ($\bm{\theta}_{global,r}, \bm{\sigma}_{global,r}^2$). Typically, the relative client sample sizes $\hat{n_i} = n_i / \sum^N n_i$ are used as weights for averaging (\cite{mcmahanCommunicationEfficientLearningDeep2017}). We retain the relative sample size $\hat{n_i}$ as a scaling factor and incorporate the inverse of the parameter variance (i.e., the parameter precision), yielding the client-specific weight $\bm{c}_i$. 
This approach is inspired by the inverse-variance aggregation scheme widely used in statistical meta-analysis, where it has been shown to yield optimal weights for combining independent random variables (\cite{borensteinIntroductionMetaanalysis2013}). All client models are initialized using He initialization (\cite{he2015delving}), with parameter variances uniformly initialized to one for both the clients and the server. The overall aggregation scheme for round $r$ is given as follows.
\begin{align}
    \bm{c}_{i,r} &=  \frac{\hat{n_i}}{\bm{\sigma}_{i,r}^2} \quad , \quad \hat{n_i} = \frac{n_i}{\sum_{i=1}^{N} n_i}\\
    \bm{\theta}_{global,r} &= \frac{\sum_{i=1}^{N} \bm{c}_{i,r} \bm{\theta}_{i,r}}{\sum_{i=1}^{N} \bm{c}_{i,r} } \\
    \quad \bm{\sigma}_{global,r}^2 &= \left(\lambda \cdot \frac{1}{\bm{\sigma}_{global, r-1}^2} + \sum_{i=1}^{N} \frac{\hat{n_i}}{\bm{\sigma}_{i,r}^2}\right)^{-1} \\
    \bm{\theta}_{i,r+1} &= \bm{\theta}_{global,r}  , \quad \bm{\sigma}_{i,r+1}^2 = \bm{\sigma}_{global,r}^2 \quad \forall i \in \{1, \dots, N\}
\end{align}


\noindent Variances are aggregated in a Bayesian update setting. To stabilize the updates, we incorporate the previous round’s global variance scaled by a forgetting factor $\lambda$. We choose \(\lambda=0.95 \) in all our experiments.

\subsection{Inference with Uncertainty Estimation}\label{sec:inference}
The central insight of FUNAvg is to leverage the predictive uncertainty for performance improvement at the inference stage. Instead of relying on Monte Carlo Dropout to obtain model samples, we leverage the global parameter variances learned during federated training.
We treat the global model as a Gaussian-distributed parameter vector and sample from this distribution by adding a perturbation $\bm{\epsilon}$ drawn from $\mathcal{N}(\bm{0}, \bm{\sigma}_{global,R}^2)$, yielding model samples $\bm{\theta}_{k}$.
\begin{equation}    
    \bm{\theta}_{k} = \bm{\theta}_{global,R} + \bm{\epsilon}, \quad \bm{\epsilon} \sim \mathcal{N}(\bm{0}, \bm{\sigma}_{global,R}^2)
\end{equation}

\noindent We perform $K$ stochastic forward passes using the sampled models  $\bm{\theta}_{k}$, computing class probabilities $\hat{\bm{p}_k}$ via the softmax function. These are then averaged to obtain the final class probabilities $\bm{\bar{p}}$ for each pixel, for each segmentation head.
\begin{equation}
    \hat{\bm{p}_k} =\text{Softmax}( f^{\theta_k}(\textbf{x})) \quad ; \quad \bm{\bar{p}} = \frac{1}{K}\sum_{k=1}^{K} \bm{\hat{p}}_k
\end{equation}
Following the approach of \cite{kwonUncertaintyQuantificationUsing2022} and \cite{tolleFUNAvgFederatedUncertainty2024}, we estimate the predictive uncertainty after $K$ sampling steps as:
\begin{equation}
 \bm{u} = \underbrace{\frac{1}{K} \left( \sum_{k=1}^{K} \mathrm{diag}(\bm{\hat{p}}_k) - \bm{\hat{p}}^{\otimes 2} \right)}_{\text{aleatoric}} + \underbrace{\frac{1}{K} \sum_{k=1}^{K} (\bm{\hat{p}}_k - \bm{\bar{p}})^{\otimes 2}}_{\text{epistemic}}   
\end{equation}
 where $\hat{\bm{p}}^{\otimes 2}$ denotes the outer product $\hat{\bm{p}} \hat{\bm{p}}^T$. This decomposition separates the data-dependent (aleatoric) and model-based (epistemic) uncertainty.
 Finally, outputs across all segmentation heads are aggregated. We reweigh the predictions from the background class using the estimated uncertainty $\bm{u}$, as proposed in FUNAvg. We refer the reader to \cite{tolleFUNAvgFederatedUncertainty2024} for further details on this step. The general methodology for our proposed approach is illustrated in Figure \ref{mainfig}.

\subsection{Experimental Setup}

\begin{table}[t]
\caption{Overview of the abdominal CT datasets used in our study. We use five datasets to serve as training clients: TotalSegmentator (TS) \cite{wasserthalTotalSegmentatorRobustSegmentation2023}, Liver Tumor Segmentation (LiTS) \cite{BILIC2023102680}, Beyond The Cranial Vault (BTCV) \cite{landman2015miccai}, AbdomenCT-1k (A1k) \cite{maAbdomenCT1KAbdominalOrgan2022}, and Learn2Reg (L2R) \cite{xuEvaluationSixRegistration2016a}. AMOS~\cite{jiAMOSLargeScaleAbdominal2022} is used as the hold-out test client. Each dataset has a different number of samples and labeled organs. The test client includes all organ classes present in the training datasets.
}
\label{tab:data}
\renewcommand{\arraystretch}{1.2}
\centering
\begin{tabular}{p{1.4cm}| p{1.27cm} p{1cm} |c c c c c c c c c c}
    \toprule
    Dataset  & \# Samples & \# Labels & DD & Eso & GB & LK & Li & Pan & RK      & Spl     & Sto     & UB \\
    \midrule
    TS   & 1139 & 7  & \cmark & \cmark & \cmark  & \cmark  & \xmark  & \xmark  & \cmark  & \xmark  & \cmark  & \cmark   \\
    LiTS & 131  & 1  & \xmark & \xmark & \xmark  & \xmark  & \cmark  & \xmark  & \xmark  & \xmark  & \xmark  & \xmark   \\
    BTCV & 30   & 6  & \xmark & \xmark & \xmark  & \cmark  & \cmark  & \cmark  & \cmark  & \cmark  & \cmark  & \xmark   \\
    A1k  & 1000 & 3  & \xmark & \xmark & \xmark  & \xmark  & \cmark  & \cmark  & \xmark  & \cmark  & \xmark  & \xmark   \\
    L2R  & 30   & 8  & \xmark & \cmark & \cmark  & \cmark  & \xmark  & \cmark  & \cmark  & \cmark  & \cmark  & \xmark   \\
    \midrule
    AMOS & 300  & 10 & \cmark & \cmark & \cmark  & \cmark  & \cmark  & \cmark  & \cmark  & \cmark  & \cmark  & \cmark   \\

    \bottomrule
\end{tabular}
\end{table}

\begin{figure}[h]
\includegraphics[width=\textwidth]{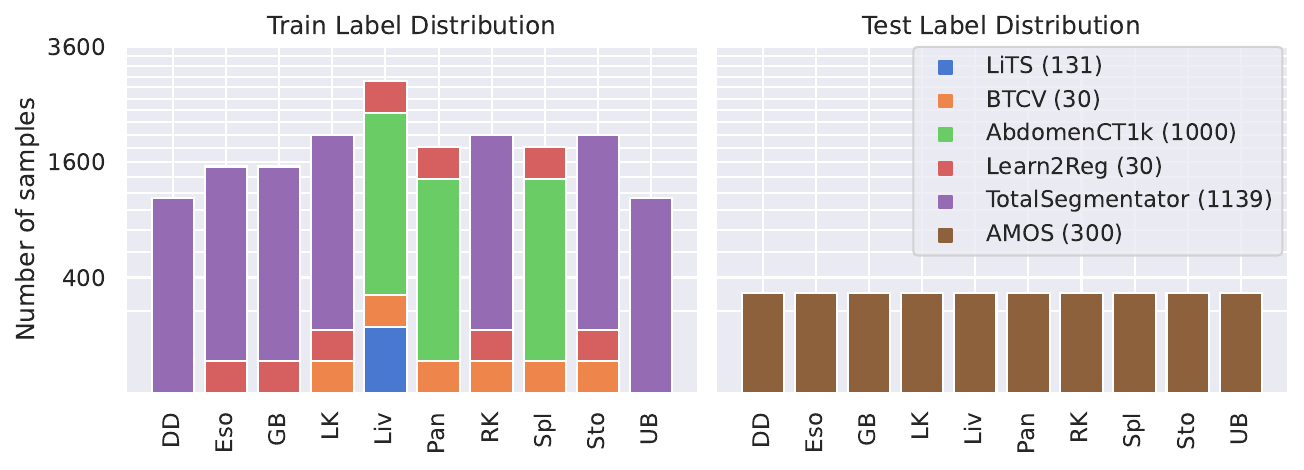}
\caption{Label distribution across the six client datasets from Table~\ref{tab:data}, illustrating the variation in organ classes and sample sizes. Each training client was assigned a subset of available organ labels such that the test client (AMOS) contains the union of all labels used during training. } \label{labeldist}
\end{figure}

\noindent We use a total of six publicly available abdominal CT datasets,  each containing a distinct subset of ten foreground organ labels\footnote{Organ labels (abbr.): Duodenum (DD), Esophagus (Eso), Gall Bladder (GB), Left Kidney (LK), Liver (Liv), Pancreas (Pan), Right Kidney (RK), Spleen (Spl), Stomach (Sto), Urinary Bladder (UB).}. We treat each dataset as an individual client in a federated learning setup, with five clients participating in training and one serving as the hold-out test client. Dataset details and label availability are summarized in Table~\ref{tab:data}. The label distribution across clients is visualized in Figure~\ref{labeldist}. Each training dataset is further split into an 80-20 train-validation split.

\noindent We use the nnUNetv2 framework (\cite{isenseeNnUNetSelfconfiguringMethod2021}) in its 2D configuration as the segmentation model. All preprocessing and architecture settings follow nnUNet’s built-in auto-configuration. The model architecture generated for the TotalSegmentator dataset is used uniformly across all clients. A fixed patch size of $256 \times 256$ is used for all clients. 
\noindent We train the federated models for 1500 rounds using SGD with momentum=0.99 and Nesterov acceleration, and a PolyLR learning rate scheduler with initial learning rate=0.01 and the polynomial exponent=0.9. The loss function is a sum of Cross-Entropy Loss and Dice Loss.

\noindent We evaluated three configurations: standalone (per-client), centralized, and federated. For the federated setup, we first compare both variants of our proposed algorithm, FIVA-G and FIVA-P (without uncertainty-weighted inference), against FedAvg. This establishes a baseline to observe the effects of inverse-variance aggregation alone.  We then evaluated both proposed variants with uncertainty-weighted inference (denoted FIVA-G+UN and FIVA-P+UN) and compared them with FUNAvg. Plotting code was adapted from~\cite{tolleFUNAvgFederatedUncertainty2024}.

\section{Results}

\begin{table}[h!]
\caption{DICE Scores ($mean_{std.}$\% from runs over five different sets of image slices) evaluated on the hold-out test sets of each of the training clients and on the hold-out test client (AMOS). The top two rows show the standalone training performance on each client's own test set (Same-client) and on the test sets of other clients (Cross-client). Centralized training is performed by jointly learning a model from all clients' data. Baseline results are first compared without uncertainty-weighted inference (FedAvg vs. FIVA-G/FIVA-P), where the best results are underlined. The next group of results incorporates uncertainty-weighted inference (FUNAvg vs. FIVA-G+UN/FIVA-P+UN), with the best results shown in bold.}\label{tab:main}
\setlength{\tabcolsep}{3.0pt}
\renewcommand{\arraystretch}{1.2}
\centering
\begin{tabular}{l|ccccc|c|c}
    \toprule
    \textbf{Method}     & \textbf{TS} & \textbf{LiTS} & \textbf{BTCV} & \textbf{A1k} & \textbf{L2R} & \textbf{AMOS} & \textbf{Mean} \\
    \midrule
    Same-client & ${87.99}_{0.27}$ & ${90.10}_{0.05}$ & ${72.50}_{0.36}$ & ${90.90}_{0.03}$ & ${70.82}_{0.30}$ & ${85.80}_{0.33}$ & ${83.02}_{0.09}$ \\
    Cross-client& ${42.89}_{1.04}$ & ${65.08}_{0.16}$ & ${64.37}_{0.41}$ & ${29.07}_{0.23}$ & ${29.33}_{0.25}$ & ${59.39}_{0.24}$ & ${48.35}_{0.21}$ \\
    \midrule
    Centralized   & ${72.42}_{0.22}$ & ${89.68}_{0.07}$ & ${62.44}_{0.54}$ & ${65.55}_{0.16}$ & ${67.03}_{0.36}$ & ${46.77}_{0.51}$ & ${67.31}_{0.10}$ \\
    \midrule
    FedAvg   & $\underline{90.72_{0.13}}$ & $57.98_{0.26}$ & $41.81_{0.59}$ & $48.54_{0.68}$ & $48.67_{1.30}$  & $37.23_{0.87}$ & $54.16_{0.18}$ \\
    FIVA-G   & $79.40_{0.18}$  & $70.67_{0.15}$ & $\underline{59.33_{0.69}}$ & $60.51_{0.08}$ & $53.48_{0.33}$ & $\underline{53.03_{0.77}}$ & $62.74_{0.12}$ \\
    FIVA-P   & ${88.02}_{0.19}$ & $\underline{{83.40}_{0.12}}$ & ${55.92}_{0.41}$ & $\underline{{61.70}_{0.15}}$ & $\underline{{56.57}_{0.25}}$ & ${47.36}_{0.89}$ & $\underline{{65.49}_{0.19}}$ \\
    \midrule
    FUNAvg   & ${\mathbf{89.02_{0.08}}}$ & $80.38_{0.29}$ & $54.57_{0.18}$ & $53.23_{0.74}$ & $52.83_{0.90}$ & $53.93_{0.60}$ & $63.99_{0.20}$ \\
    FIVA-G+UN  & $78.08_{0.20}$ & $81.50_{0.15}$  & $60.76_{0.72}$ & $60.66_{0.07}$ & $54.33_{0.22}$ & $\mathbf{55.53_{0.68}}$ & $65.14_{0.13}$ \\
    FIVA-P+UN  & ${87.03}_{0.17}$ & $\mathbf{{86.45}_{0.10}}$ & $\mathbf{{56.97}_{0.53}}$ & $\mathbf{{61.92}_{0.14}}$ & $\mathbf{{57.03}_{0.28}}$ & ${49.15}_{0.82}$ & $\mathbf{{66.42}_{0.21}}$ \\
\hline
    \bottomrule
\end{tabular}

\end{table}

\begin{figure}[h!]
\centering
\includegraphics[width=\textwidth]{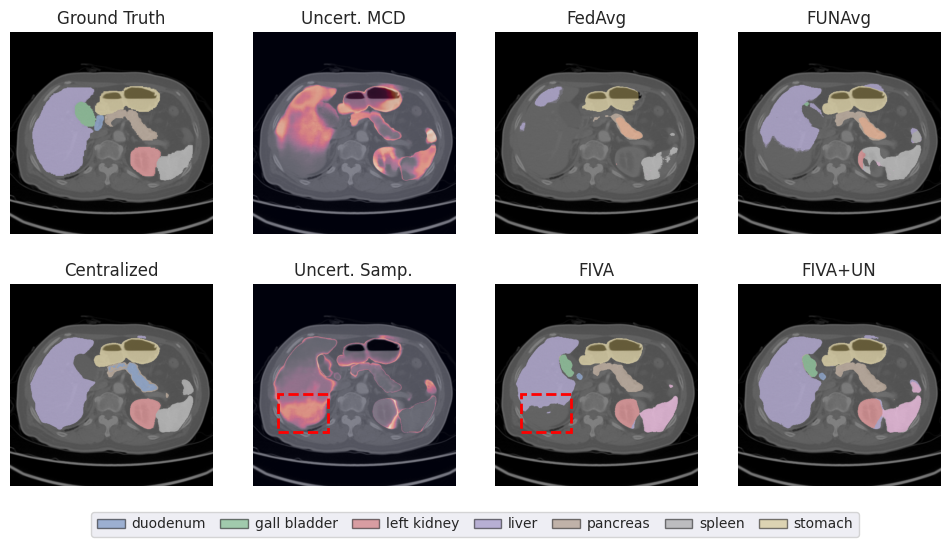}
\caption{Qualitative results on the AMOS test set highlight the core operating principle of uncertainty weighting. Sampling-based uncertainty (Uncert. Samp.) captures uncertainties near organ boundaries and in partially segmented regions. In contrast, MC Dropout-based uncertainty (Uncert. MCD) fails to capture the lower regions of the liver (highlighted in red). The last column shows false background predictions suppressed in areas of high uncertainty.}
\label{qual}
\end{figure}

\noindent Table \ref{tab:main} summarizes the performance of our method compared to the baselines across training clients and the held-out AMOS test client. Both variants, FIVA-G and FIVA-P, improve on FedAvg for all but one client, demonstrating that inverse-variance aggregation offers a standalone improvement during training. The parameter-based variant of our proposed method, FIVA-P, improves over FedAvg by nearly 11 percentage points in mean Dice score (54.16 vs. 65.49).  When combined with the sampling-based uncertainty weighting strategy (FIVA-G+UN and FIVA-P+UN), we observe similar improvements as compared to FUNAvg, which uses MC Dropout-based uncertainty. We also observe that uncertainty-weighted inference generally improves performance over naive inference, consistent with the findings of \cite{tolleFUNAvgFederatedUncertainty2024}.

\begin{figure}[h!]
  \centering

  \begin{minipage}[b]{0.48\textwidth}
    \centering
    \includegraphics[width=\linewidth]{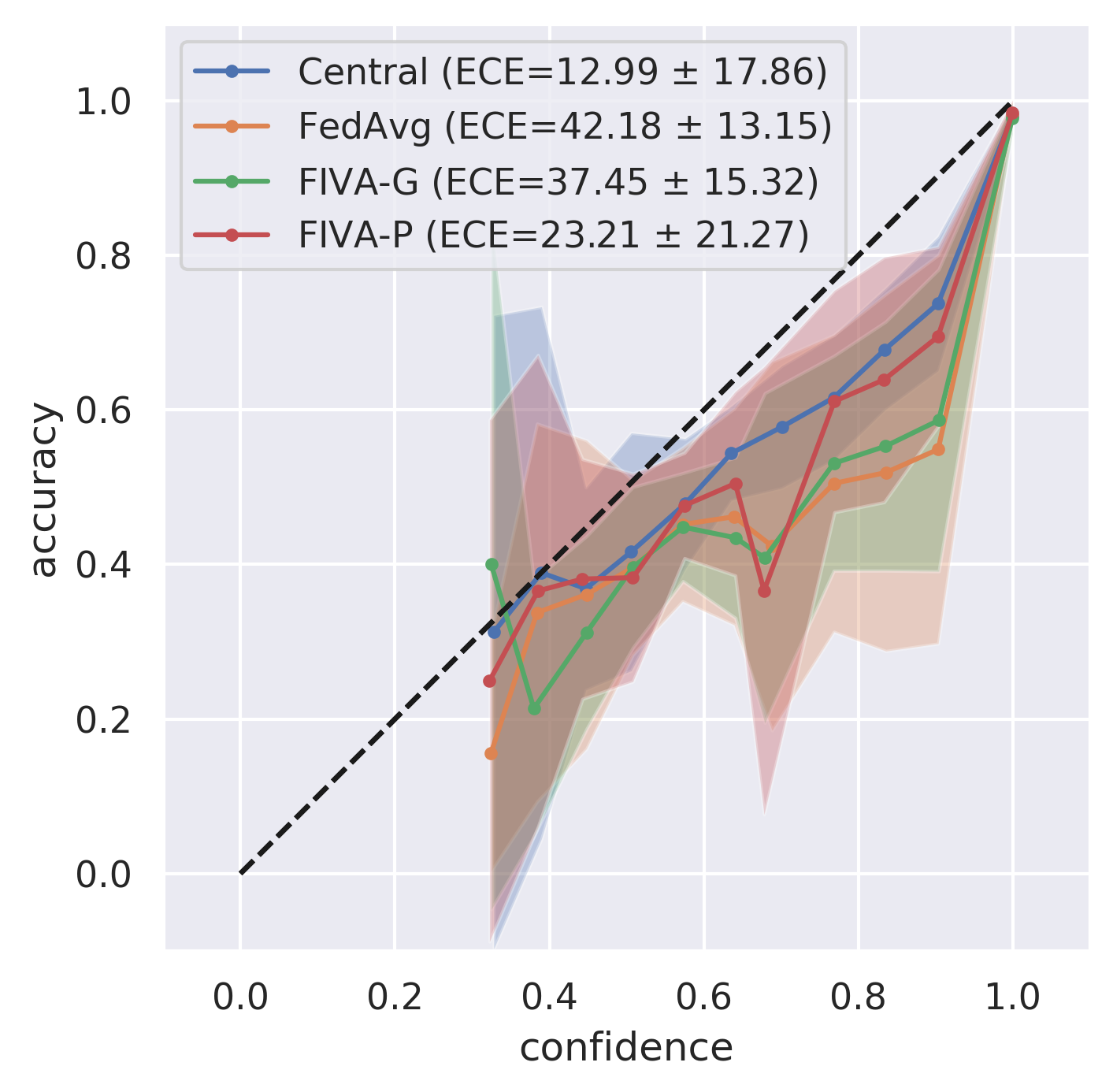}
    \par
    \parbox{0.9\linewidth}{\centering (a) Model calibration \textbf{without} uncertainty weighted inference.}
  \end{minipage}
  \hfill
  \begin{minipage}[b]{0.48\textwidth}
    \centering
    \includegraphics[width=\linewidth]{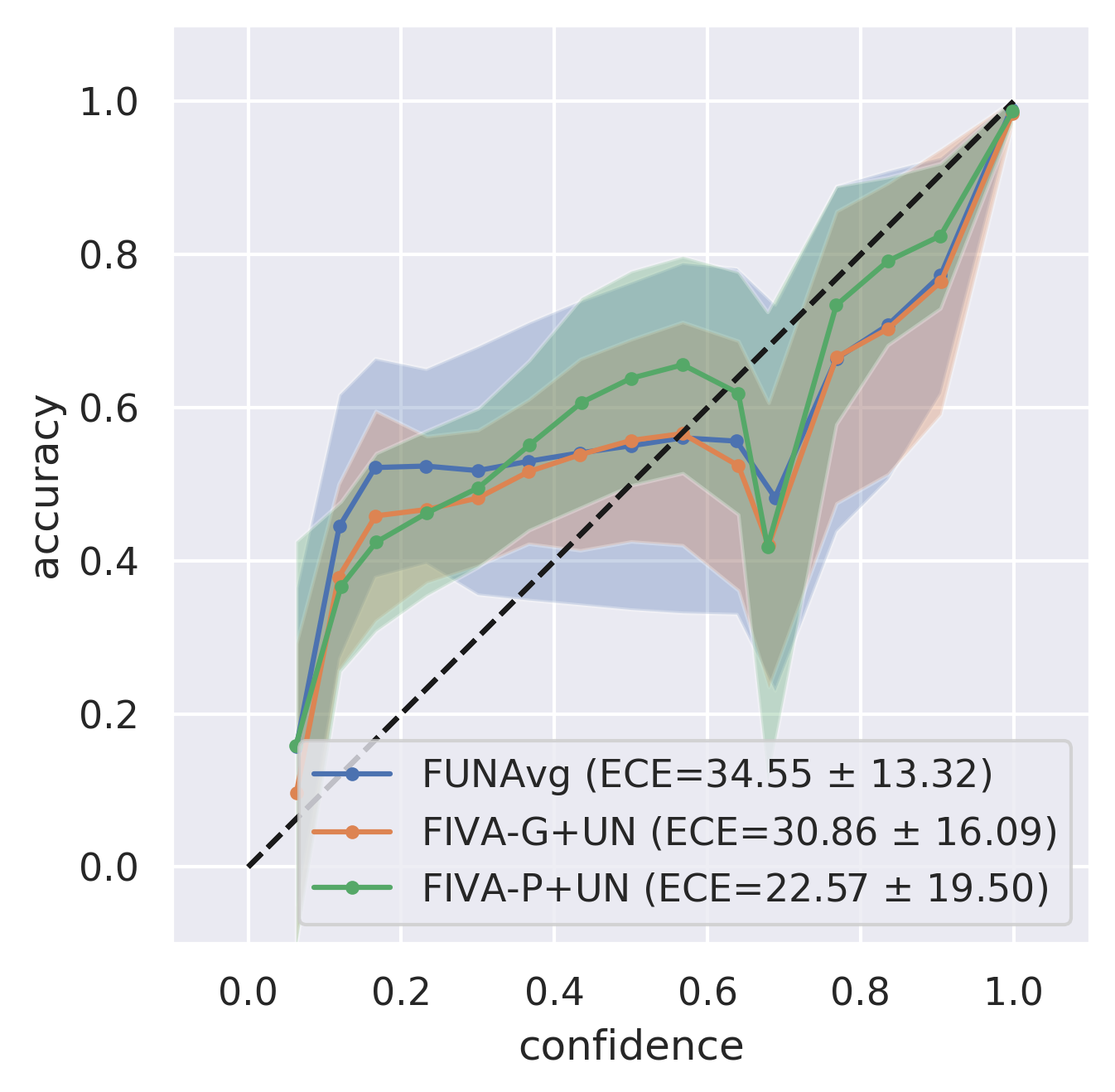}
    \centering
    \par
    \parbox{0.9\linewidth}{ (b) Model calibration \textbf{with} uncertainty weighted inference.}
  \end{minipage}

  \caption{Reliability diagrams for the tested methods with the label-wise aggregated ECE (mean $\pm$ std). FIVA and FIVA+UN show improvement in calibration error over the baselines. Notably, uncertainty-weighted inference (b) leads to a more balanced model calibration compared to naive averaging (a).}
  \label{fig:calib}
\end{figure}

\noindent The qualitative results in Figure \ref{qual} further illustrate the benefit of uncertainty-driven inference. FIVA+UN captures subtle boundary ambiguities that are overlooked by the MC dropout-based method, especially in anatomically complex regions or small organs. Additional qualitative results on the AMOS test set are shown in Figure \ref{app:qual_fig} in Appendix \ref{app:additional}.
\noindent Reliability diagrams (Figure \ref{fig:calib}) show that FIVA models exhibit better-calibrated predictions with lower Expected Calibration Error (ECE) compared to FedAvg and FUNAvg. These results establish FIVA’s accuracy and calibration reliability in federated segmentation under real-world heterogeneity.

\section{Discussion} 

The results we obtained highlight some important takeaways. First, we propose a sampling-based method for modeling uncertainty that improves both the training and inference stages in real-world federated learning.
Second, we observe consistent improvements in Dice scores when using sampling-based uncertainty estimates compared to MC Dropout. Although MC Dropout is a widely used approach for predictive uncertainty, it reduces model capacity during inference by randomly disabling neurons, potentially removing those most salient for segmentation. In contrast, our sampling-based approach preserves full model capacity as long as parameter variances remain bounded.
Third, qualitative results show that the uncertainty estimates derived from our sampling method correctly emphasize organ boundaries, regions that naturally exhibit high uncertainty. Finally, our results reinforce earlier findings from FUNAvg: uncertainty-weighted logit averaging at inference time helps suppress false-positive background predictions, improving segmentation accuracy.

\noindent Although the proposed approach shows an improvement in segmentation performance, it incurs an additional computation cost. Accounting for the added overhead of estimating the variances and uncertainties would be crucial for deployment in compute-limited environments. We briefly discuss this overhead below.

\paragraph{Time complexity.}Let the model have $M$ trainable parameters, and assume each input data point is $D$-dimensional. For a total of $T$ mini-batch SGD updates using batches of size $B$, the computational time complexity of standard training is $\mathcal{O}(TMBD)$, accounting for both forward and backward passes. Our proposed method, FIVA, introduces two additional vectorized operations per update: one for updating the running mean of gradients/parameters and another for computing the sum of squared errors, both using Welford’s algorithm. These operations introduce an added computational cost of $\mathcal{O}(TM)$. Given that $M \ll MBD$, and noting that additional computations are parallelizable with a GPU, the asymptotic complexity remains $\mathcal{O}(TMBD)$. Therefore, the runtime overhead introduced by FIVA is negligible in practical settings, particularly when training deep models on large datasets.

\paragraph{Space complexity.} In standard training, each client typically maintains three tensors of size $\mathcal{O}(M)$: model parameters, gradient estimates, and optimizer states such as momentum, leading to a total space complexity of $\mathcal{O}(3M)$. FIVA augments this with three additional $\mathcal{O}(M)$ tensors per client to store the running means, the sum of squared errors, and the variances of the gradients/parameters. This results in a total space complexity of $\mathcal{O}(6M)$ per client. Although FIVA requires double the memory compared to standard training, it is significantly more memory efficient than storing full gradient/parameter histories that require $\mathcal{O}(TM)$ space, where $T$ is the number of training iterations. This is due to Welford’s algorithm, which uses constant memory per parameter for the variance estimation.


\paragraph{Limitations.}
While the inverse-variance aggregation scheme is well grounded in statistical meta-analysis, its use in machine learning remains relatively unexplored. For this method to function reliably, parameter variances must remain upper- and lower-bounded throughout the training. In low-data regimes, variance estimates can become unreliable, potentially destabilizing training.
Although our approach improves upon existing baselines, a significant gap remains between federated and standalone performance, limiting immediate clinical applicability. Furthermore, calibration errors, while improved, are still relatively high compared to typical values reported in classical machine learning settings.

\noindent Overall, this work bridges two important paradigms in machine learning for healthcare: uncertainty estimation and federated learning, demonstrating how insights from one can meaningfully improve the other.
We hope that future research will focus on variance stabilization techniques, improved calibration under heterogeneity, and more efficient approximations of uncertainty to make these methods viable for broader adoption.

\bibliography{references}

\begin{thebibliography}{35}
\providecommand{\natexlab}[1]{#1}
\providecommand{\url}[1]{\texttt{#1}}
\expandafter\ifx\csname urlstyle\endcsname\relax
  \providecommand{\doi}[1]{doi: #1}\else
  \providecommand{\doi}{doi: \begingroup \urlstyle{rm}\Url}\fi

\bibitem[Asokan et~al.(2024)Asokan, Benjamin, Yaqub, and
  Nandakumar]{asokan2024federated}
Mothilal Asokan, Joseph~Geo Benjamin, Mohammad Yaqub, and Karthik Nandakumar.
\newblock A federated learning-friendly approach for parameter-efficient
  fine-tuning of sam in 3d segmentation.
\newblock In \emph{International Conference on Medical Image Computing and
  Computer-Assisted Intervention}, pages 226--235. Springer, 2024.

\bibitem[Baumgartner et~al.(2019)Baumgartner, Tezcan, Chaitanya, H{\"o}tker,
  Muehlematter, Schawkat, Becker, Donati, and
  Konukoglu]{baumgartnerPHiSegCapturingUncertainty2019}
Christian~F. Baumgartner, Kerem~C. Tezcan, Krishna Chaitanya, Andreas~M.
  H{\"o}tker, Urs~J. Muehlematter, Khoschy Schawkat, Anton~S. Becker, Olivio
  Donati, and Ender Konukoglu.
\newblock {{PHiSeg}}: {{Capturing Uncertainty}} in {{Medical Image
  Segmentation}}, July 2019.

\bibitem[Bernecker et~al.(2022)Bernecker, Peters, Schlett, Bamberg, Theis,
  Rueckert, Wei{\ss}, and Albarqouni]{bernecker2022fednorm}
Tobias Bernecker, Annette Peters, Christopher~L Schlett, Fabian Bamberg, Fabian
  Theis, Daniel Rueckert, Jakob Wei{\ss}, and Shadi Albarqouni.
\newblock Fednorm: Modality-based normalization in federated learning for
  multi-modal liver segmentation.
\newblock \emph{arXiv preprint arXiv:2205.11096}, 2022.

\bibitem[Bilic et~al.(2023)Bilic, Christ, Li, Vorontsov, Ben-Cohen, Kaissis,
  Szeskin, Jacobs, Mamani, Chartrand, Lohöfer, Holch, Sommer, Hofmann,
  Hostettler, Lev-Cohain, Drozdzal, Amitai, Vivanti, Sosna, Ezhov, Sekuboyina,
  Navarro, Kofler, Paetzold, Shit, Hu, Lipková, Rempfler, Piraud, Kirschke,
  Wiestler, Zhang, Hülsemeyer, Beetz, Ettlinger, Antonelli, Bae, Bellver, Bi,
  Chen, Chlebus, Dam, Dou, Fu, Georgescu, i~Nieto, Gruen, Han, Heng, Hesser,
  Moltz, Igel, Isensee, Jäger, Jia, Kaluva, Khened, Kim, Kim, Kim, Kohl,
  Konopczynski, Kori, Krishnamurthi, Li, Li, Li, Li, Lowengrub, Ma, Maier-Hein,
  Maninis, Meine, Merhof, Pai, Perslev, Petersen, Pont-Tuset, Qi, Qi, Rippel,
  Roth, Sarasua, Schenk, Shen, Torres, Wachinger, Wang, Weninger, Wu, Xu, Yang,
  Yu, Yuan, Yue, Zhang, Cardoso, Bakas, Braren, Heinemann, Pal, Tang, Kadoury,
  Soler, {van Ginneken}, Greenspan, Joskowicz, and Menze]{BILIC2023102680}
Patrick Bilic, Patrick Christ, Hongwei~Bran Li, Eugene Vorontsov, Avi
  Ben-Cohen, Georgios Kaissis, Adi Szeskin, Colin Jacobs, Gabriel
  Efrain~Humpire Mamani, Gabriel Chartrand, Fabian Lohöfer, Julian~Walter
  Holch, Wieland Sommer, Felix Hofmann, Alexandre Hostettler, Naama Lev-Cohain,
  Michal Drozdzal, Michal~Marianne Amitai, Refael Vivanti, Jacob Sosna, Ivan
  Ezhov, Anjany Sekuboyina, Fernando Navarro, Florian Kofler, Johannes~C.
  Paetzold, Suprosanna Shit, Xiaobin Hu, Jana Lipková, Markus Rempfler, Marie
  Piraud, Jan Kirschke, Benedikt Wiestler, Zhiheng Zhang, Christian
  Hülsemeyer, Marcel Beetz, Florian Ettlinger, Michela Antonelli, Woong Bae,
  Míriam Bellver, Lei Bi, Hao Chen, Grzegorz Chlebus, Erik~B. Dam, Qi~Dou,
  Chi-Wing Fu, Bogdan Georgescu, Xavier~Giró i~Nieto, Felix Gruen, Xu~Han,
  Pheng-Ann Heng, Jürgen Hesser, Jan~Hendrik Moltz, Christian Igel, Fabian
  Isensee, Paul Jäger, Fucang Jia, Krishna~Chaitanya Kaluva, Mahendra Khened,
  Ildoo Kim, Jae-Hun Kim, Sungwoong Kim, Simon Kohl, Tomasz Konopczynski,
  Avinash Kori, Ganapathy Krishnamurthi, Fan Li, Hongchao Li, Junbo Li,
  Xiaomeng Li, John Lowengrub, Jun Ma, Klaus Maier-Hein, Kevis-Kokitsi Maninis,
  Hans Meine, Dorit Merhof, Akshay Pai, Mathias Perslev, Jens Petersen, Jordi
  Pont-Tuset, Jin Qi, Xiaojuan Qi, Oliver Rippel, Karsten Roth, Ignacio
  Sarasua, Andrea Schenk, Zengming Shen, Jordi Torres, Christian Wachinger,
  Chunliang Wang, Leon Weninger, Jianrong Wu, Daguang Xu, Xiaoping Yang, Simon
  Chun-Ho Yu, Yading Yuan, Miao Yue, Liping Zhang, Jorge Cardoso, Spyridon
  Bakas, Rickmer Braren, Volker Heinemann, Christopher Pal, An~Tang, Samuel
  Kadoury, Luc Soler, Bram {van Ginneken}, Hayit Greenspan, Leo Joskowicz, and
  Bjoern Menze.
\newblock The liver tumor segmentation benchmark (lits).
\newblock \emph{Medical Image Analysis}, 84:\penalty0 102680, 2023.
\newblock ISSN 1361-8415.
\newblock \doi{https://doi.org/10.1016/j.media.2022.102680}.
\newblock URL
  \url{https://www.sciencedirect.com/science/article/pii/S1361841522003085}.

\bibitem[Borenstein(2013)]{borensteinIntroductionMetaanalysis2013}
Michael Borenstein, editor.
\newblock \emph{Introduction to Meta-Analysis}.
\newblock Wiley, Chichester, nachdr. edition, 2013.
\newblock ISBN 978-0-470-05724-7.

\bibitem[Butoi et~al.(2023)Butoi, Ortiz, Ma, Sabuncu, Guttag, and
  Dalca]{butoi2023universeg}
Victor~Ion Butoi, Jose Javier~Gonzalez Ortiz, Tianyu Ma, Mert~R Sabuncu, John
  Guttag, and Adrian~V Dalca.
\newblock Universeg: Universal medical image segmentation.
\newblock In \emph{Proceedings of the IEEE/CVF International Conference on
  Computer Vision}, pages 21438--21451, 2023.

\bibitem[Chen et~al.(2019)Chen, Ma, and Zheng]{chen2019med3d}
Sihong Chen, Kai Ma, and Yefeng Zheng.
\newblock Med3d: Transfer learning for 3d medical image analysis.
\newblock \emph{arXiv preprint arXiv:1904.00625}, 2019.

\bibitem[Gal and Ghahramani(2016)]{gal2016dropout}
Yarin Gal and Zoubin Ghahramani.
\newblock Dropout as a bayesian approximation: Representing model uncertainty
  in deep learning.
\newblock In \emph{international conference on machine learning}, pages
  1050--1059. PMLR, 2016.

\bibitem[Gao et~al.(2024)Gao, Li, Liu, Zhou, Zhang, and
  Metaxas]{gaoTrainingMedicalResident2024}
Yunhe Gao, Zhuowei Li, Di~Liu, Mu~Zhou, Shaoting Zhang, and Dimitris~N.
  Metaxas.
\newblock Training {{Like}} a {{Medical Resident}}: {{Context-Prior Learning
  Toward Universal Medical Image Segmentation}}, April 2024.

\bibitem[Hartung(2008)]{hartungStatisticalMetaanalysisApplications2008}
Joachim Hartung.
\newblock \emph{Statistical Meta-Analysis with Applications}.
\newblock Wiley, Hoboken, NJ, 2008.
\newblock ISBN 978-0-470-29089-7 978-0-470-38633-0.

\bibitem[He et~al.(2015)He, Zhang, Ren, and Sun]{he2015delving}
Kaiming He, Xiangyu Zhang, Shaoqing Ren, and Jian Sun.
\newblock Delving deep into rectifiers: Surpassing human-level performance on
  imagenet classification.
\newblock In \emph{Proceedings of the IEEE international conference on computer
  vision}, pages 1026--1034, 2015.

\bibitem[Isensee et~al.(2021)Isensee, Jaeger, Kohl, Petersen, and
  {Maier-Hein}]{isenseeNnUNetSelfconfiguringMethod2021}
Fabian Isensee, Paul~F. Jaeger, Simon A.~A. Kohl, Jens Petersen, and Klaus~H.
  {Maier-Hein}.
\newblock {{nnU-Net}}: A self-configuring method for deep learning-based
  biomedical image segmentation.
\newblock \emph{Nature Methods}, 18\penalty0 (2):\penalty0 203--211, February
  2021.
\newblock ISSN 1548-7091, 1548-7105.
\newblock \doi{10.1038/s41592-020-01008-z}.

\bibitem[Jalal et~al.(2024)Jalal, {\'S}liwi{\'n}ska, Wojciechowski,
  Kucyba{\l}a, Rozynek, Krupa, Matusik, Jarczewski, and
  Tabor]{jalalEvaluatingUncertaintyQuantification2024}
Nyaz Jalal, Ma{\l}gorzata {\'S}liwi{\'n}ska, Wadim Wojciechowski, Iwona
  Kucyba{\l}a, Mi{\l}osz Rozynek, Kamil Krupa, Patrycja Matusik, Jaros{\l}aw
  Jarczewski, and Zbis{\l}aw Tabor.
\newblock Evaluating {{Uncertainty Quantification}} in {{Medical Image
  Segmentation}}: {{A Multi-Dataset}}, {{Multi-Algorithm Study}}.
\newblock \emph{Applied Sciences}, 14\penalty0 (21):\penalty0 10020, November
  2024.
\newblock ISSN 2076-3417.
\newblock \doi{10.3390/app142110020}.

\bibitem[Ji et~al.(2022)Ji, Bai, Ge, Yang, Zhu, Zhang, Li, Zhanng, Ma, Wan, and
  Luo]{jiAMOSLargeScaleAbdominal2022}
Yuanfeng Ji, Haotian Bai, Chongjian Ge, Jie Yang, Ye~Zhu, Ruimao Zhang, Zhen
  Li, Lingyan Zhanng, Wanling Ma, Xiang Wan, and Ping Luo.
\newblock {{AMOS}}: {{A Large-Scale Abdominal Multi-Organ Benchmark}} for
  {{Versatile Medical Image Segmentation}}.
\newblock \emph{Advances in Neural Information Processing Systems},
  35:\penalty0 36722--36732, December 2022.

\bibitem[Jiang et~al.(2023)Jiang, Roth, Li, Yang, Zhao, Nath, Xu, Dou, and
  Xu]{jiangFairFederatedMedical2023a}
Meirui Jiang, Holger~R Roth, Wenqi Li, Dong Yang, Can Zhao, Vishwesh Nath,
  Daguang Xu, Qi~Dou, and Ziyue Xu.
\newblock Fair {{Federated Medical Image Segmentation}} via {{Client
  Contribution Estimation}}.
\newblock \emph{CVPR}, 2023.

\bibitem[Judge et~al.(2022)Judge, Bernard, Porumb, Chartsias, Beqiri, and
  Jodoin]{judgeCRISPReliableUncertainty2022}
Thierry Judge, Olivier Bernard, Mihaela Porumb, Agis Chartsias, Arian Beqiri,
  and Pierre-Marc Jodoin.
\newblock {{CRISP}} - {{Reliable Uncertainty Estimation}} for {{Medical Image
  Segmentation}}, June 2022.

\bibitem[Kendall and Gal(2017)]{kendallWhatUncertaintiesWe2017}
Alex Kendall and Yarin Gal.
\newblock What {{Uncertainties Do We Need}} in {{Bayesian Deep Learning}} for
  {{Computer Vision}}?, October 2017.

\bibitem[Koutsoubis et~al.(2024)Koutsoubis, Yilmaz, Ramachandran, Schabath, and
  Rasool]{koutsoubis2024privacy}
Nikolas Koutsoubis, Yasin Yilmaz, Ravi~P Ramachandran, Matthew Schabath, and
  Ghulam Rasool.
\newblock Privacy preserving federated learning in medical imaging with
  uncertainty estimation.
\newblock \emph{arXiv preprint arXiv:2406.12815}, 2024.

\bibitem[Kwon et~al.(2022)Kwon, Won, Kim, and
  Paik]{kwonUncertaintyQuantificationUsing2022}
Yongchan Kwon, Joong-Ho Won, Beom~Joon Kim, and Myunghee~Cho Paik.
\newblock Uncertainty quantification using {{Bayesian}} neural networks in
  classification: {{Application}} to ischemic stroke lesion segmentation.
\newblock In \emph{Medical {{Imaging}} with {{Deep Learning}}}, July 2022.

\bibitem[Lakshminarayanan et~al.(2017)Lakshminarayanan, Pritzel, and
  Blundell]{lakshminarayananSimpleScalablePredictive2017a}
Balaji Lakshminarayanan, Alexander Pritzel, and Charles Blundell.
\newblock Simple and {{Scalable Predictive Uncertainty Estimation}} using
  {{Deep Ensembles}}.
\newblock In \emph{Advances in {{Neural Information Processing Systems}}},
  volume~30. Curran Associates, Inc., 2017.

\bibitem[Landman et~al.(2015)Landman, Xu, Igelsias, Styner, Langerak, and
  Klein]{landman2015miccai}
Bennett Landman, Zhoubing Xu, Juan Igelsias, Martin Styner, Thomas Langerak,
  and Arno Klein.
\newblock Miccai multi-atlas labeling beyond the cranial vault--workshop and
  challenge.
\newblock In \emph{Proc. MICCAI multi-atlas labeling beyond cranial
  vault—workshop challenge}, volume~5, page~12. Munich, Germany, 2015.

\bibitem[Li et~al.(2020)Li, Sahu, Zaheer, Sanjabi, Talwalkar, and
  Smith]{liFederatedOptimizationHeterogeneous2020a}
Tian Li, Anit~Kumar Sahu, Manzil Zaheer, Maziar Sanjabi, Ameet Talwalkar, and
  Virginia Smith.
\newblock Federated {{Optimization}} in {{Heterogeneous Networks}}.
\newblock \emph{Proceedings of Machine Learning and Systems}, 2:\penalty0
  429--450, March 2020.

\bibitem[Ma et~al.(2022)Ma, Zhang, Gu, Zhu, Ge, Zhang, An, Wang, Wang, Liu,
  Cao, Zhang, Liu, Wang, Li, He, and Yang]{maAbdomenCT1KAbdominalOrgan2022}
Jun Ma, Yao Zhang, Song Gu, Cheng Zhu, Cheng Ge, Yichi Zhang, Xingle An,
  Congcong Wang, Qiyuan Wang, Xin Liu, Shucheng Cao, Qi~Zhang, Shangqing Liu,
  Yunpeng Wang, Yuhui Li, Jian He, and Xiaoping Yang.
\newblock {{AbdomenCT-1K}}: {{Is Abdominal Organ Segmentation}} a {{Solved
  Problem}}?
\newblock \emph{IEEE Transactions on Pattern Analysis and Machine
  Intelligence}, 44\penalty0 (10):\penalty0 6695--6714, October 2022.
\newblock ISSN 0162-8828, 2160-9292, 1939-3539.
\newblock \doi{10.1109/TPAMI.2021.3100536}.

\bibitem[McMahan et~al.(2017)McMahan, Moore, Ramage, Hampson, and
  y~Arcas]{mcmahanCommunicationEfficientLearningDeep2017}
Brendan McMahan, Eider Moore, Daniel Ramage, Seth Hampson, and Blaise~Aguera
  y~Arcas.
\newblock Communication-{{Efficient Learning}} of {{Deep Networks}} from
  {{Decentralized Data}}.
\newblock In \emph{Proceedings of the 20th {{International Conference}} on
  {{Artificial Intelligence}} and {{Statistics}}}, pages 1273--1282. PMLR,
  April 2017.

\bibitem[Rauniyar et~al.(2024)Rauniyar, Hagos, Jha, H{\aa}keg{\aa}rd, Bagci,
  Rawat, and Vlassov]{rauniyarFederatedLearningMedical2024}
Ashish Rauniyar, Desta~Haileselassie Hagos, Debesh Jha, Jan~Erik
  H{\aa}keg{\aa}rd, Ulas Bagci, Danda~B. Rawat, and Vladimir Vlassov.
\newblock Federated {{Learning}} for {{Medical Applications}}: {{A Taxonomy}},
  {{Current Trends}}, {{Challenges}}, and {{Future Research Directions}}.
\newblock \emph{IEEE Internet of Things Journal}, 11\penalty0 (5):\penalty0
  7374--7398, March 2024.
\newblock ISSN 2327-4662, 2372-2541.
\newblock \doi{10.1109/JIOT.2023.3329061}.

\bibitem[Ronneberger et~al.(2015)Ronneberger, Fischer, and
  Brox]{ronneberger2015u}
Olaf Ronneberger, Philipp Fischer, and Thomas Brox.
\newblock U-net: Convolutional networks for biomedical image segmentation.
\newblock In \emph{Medical image computing and computer-assisted
  intervention--MICCAI 2015: 18th international conference, Munich, Germany,
  October 5-9, 2015, proceedings, part III 18}, pages 234--241. Springer, 2015.

\bibitem[T{\"o}lle et~al.(2024)T{\"o}lle, Navarro, Eble, Wolf, Menze, and
  Engelhardt]{tolleFUNAvgFederatedUncertainty2024}
Malte T{\"o}lle, Fernando Navarro, Sebastian Eble, Ivo Wolf, Bjoern Menze, and
  Sandy Engelhardt.
\newblock {{FUNAvg}}: {{Federated Uncertainty Weighted Averaging}} for
  {{Datasets}} with {{Diverse Labels}}, July 2024.

\bibitem[Ulrich et~al.(2023)Ulrich, Isensee, Wald, Zenk, Baumgartner, and
  {Maier-Hein}]{ulrichMultiTalentMultiDatasetApproach2023}
Constantin Ulrich, Fabian Isensee, Tassilo Wald, Maximilian Zenk, Michael
  Baumgartner, and Klaus~H. {Maier-Hein}.
\newblock {{MultiTalent}}: {{A Multi-Dataset Approach}} to {{Medical Image
  Segmentation}}.
\newblock volume 14222, pages 648--658. 2023.
\newblock \doi{10.1007/978-3-031-43898-1_62}.

\bibitem[Wasserthal et~al.(2023)Wasserthal, Breit, Meyer, Pradella, Hinck,
  Sauter, Heye, Boll, Cyriac, Yang, Bach, and
  Segeroth]{wasserthalTotalSegmentatorRobustSegmentation2023}
Jakob Wasserthal, Hanns-Christian Breit, Manfred~T. Meyer, Maurice Pradella,
  Daniel Hinck, Alexander~W. Sauter, Tobias Heye, Daniel~T. Boll, Joshy Cyriac,
  Shan Yang, Michael Bach, and Martin Segeroth.
\newblock {{TotalSegmentator}}: {{Robust Segmentation}} of 104 {{Anatomic
  Structures}} in {{CT Images}}.
\newblock \emph{Radiology: Artificial Intelligence}, 5\penalty0 (5):\penalty0
  e230024, September 2023.
\newblock ISSN 2638-6100.
\newblock \doi{10.1148/ryai.230024}.

\bibitem[Welford(1962)]{Welford01081962}
B.~P. Welford.
\newblock Note on a method for calculating corrected sums of squares and
  products.
\newblock \emph{Technometrics : a journal of statistics for the physical,
  chemical, and engineering sciences}, 4\penalty0 (3):\penalty0 419--420, 1962.
\newblock \doi{10.1080/00401706.1962.10490022}.

\bibitem[Xu et~al.(2023)Xu, Deng, Gateno, and
  Yan]{xuFederatedMultiOrganSegmentation2023}
Xuanang Xu, Hannah~H. Deng, Jamie Gateno, and Pingkun Yan.
\newblock Federated {{Multi-Organ Segmentation With Inconsistent Labels}}.
\newblock \emph{IEEE Transactions on Medical Imaging}, 42\penalty0
  (10):\penalty0 2948--2960, October 2023.
\newblock ISSN 0278-0062, 1558-254X.
\newblock \doi{10.1109/TMI.2023.3270140}.

\bibitem[Xu et~al.(2024)Xu, Deng, Chen, Kuang, Barber, Kim, Gateno, Xia, and
  Yan]{xu2024federated}
Xuanang Xu, Hannah~H Deng, Tianyi Chen, Tianshu Kuang, Joshua~C Barber,
  Daeseung Kim, Jaime Gateno, James~J Xia, and Pingkun Yan.
\newblock Federated cross learning for medical image segmentation.
\newblock In \emph{Medical Imaging with Deep Learning}, pages 1441--1452. PMLR,
  2024.

\bibitem[Xu et~al.(2016)Xu, Lee, Heinrich, Modat, Rueckert, Ourselin, Abramson,
  and Landman]{xuEvaluationSixRegistration2016a}
Zhoubing Xu, Christopher~P. Lee, Mattias~P. Heinrich, Marc Modat, Daniel
  Rueckert, Sebastien Ourselin, Richard~G. Abramson, and Bennett~A. Landman.
\newblock Evaluation of {{Six Registration Methods}} for the {{Human Abdomen}}
  on {{Clinically Acquired CT}}.
\newblock \emph{IEEE Transactions on Biomedical Engineering}, 63\penalty0
  (8):\penalty0 1563--1572, August 2016.
\newblock ISSN 0018-9294, 1558-2531.
\newblock \doi{10.1109/TBME.2016.2574816}.

\bibitem[Zhang et~al.(2022)Zhang, Li, Li, Guo, and
  Shao]{zhangPersonalizedFederatedLearning2022}
Xu~Zhang, Yinchuan Li, Wenpeng Li, Kaiyang Guo, and Yunfeng Shao.
\newblock Personalized {{Federated Learning}} via {{Variational Bayesian
  Inference}}.
\newblock In \emph{Proceedings of the 39th {{International Conference}} on
  {{Machine Learning}}}, pages 26293--26310. PMLR, June 2022.

\bibitem[Zhang et~al.(2023)Zhang, Ghosh, Xia, and
  Mascolo]{zhangUncertaintyQuantificationFederated2023}
Yuwei Zhang, Abhirup Ghosh, Tong Xia, and Cecilia Mascolo.
\newblock Uncertainty {{Quantification}} in {{Federated Learning}} for
  {{Heterogeneous Health Data}}.
\newblock 2023.

\end{thebibliography}
\newpage
\appendix
\section{Additional Qualitative Results}
\label{app:additional}
\begin{figure}[h!]
\centering
\includegraphics[width=\textwidth]{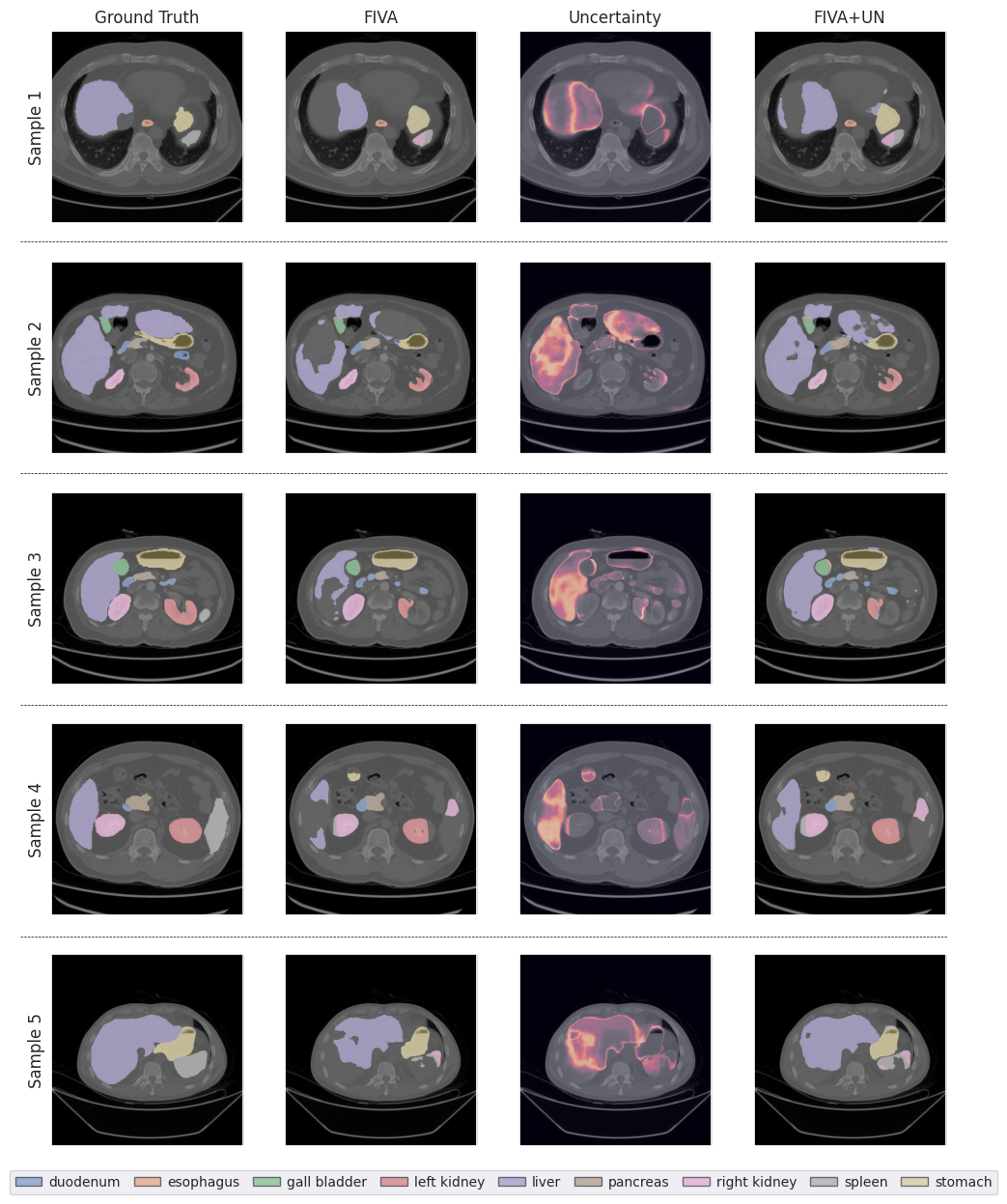}
\caption{Additional qualitative results on different samples from the AMOS test set.}
\label{app:qual_fig}
\end{figure}
\end{document}